\documentclass[%
 amsmath,amssymb,
]{revtex4-2}

\usepackage{graphicx}
\usepackage{dcolumn}
\usepackage{comment}
\usepackage{gensymb}
\usepackage{bm}
\usepackage{hyperref}

\usepackage[
total={6.5in,8.75in}, top=1.2in, left=0.9in, includefoot,
]{geometry}

\begin{document}

\preprint{APS/123-QED}

\title{Controlled and impulsive compression of an entrapped air bubble during impact}

\author{Utkarsh Jain}
\author{Devaraj van der Meer}%
\affiliation{%
Physics of Fluids Group and Max Planck Center Twente for Complex Fluid Dynamics, MESA+ Institute and J. M. Burgers Centre for Fluid Dynamics, University of Twente\\ Enschede, The Netherlands
}%

\begin{abstract}
Wave slamming onto a structure is often accompanied by the entrapment of an air pocket. A large scale impact typically has a rapidly evolving and disturbed liquid-gas interface, such that several bubbles are entrapped upon impact. While it is largely understood how the peak pressure is created by liquid coming into contact with the solid structure, it is more challenging to ascertain how an isolated air pocket is pressurised by an impulsive impact, and how the maximum impact pressure inside this bubble evolves. We study such a Bagnold-type impulsive compression of an air bubble by performing well-controlled experiments, where we use an inverted, hollow cone as an impactor. The cone is kept immersed throughout in a water bath, such that it encloses an air bubble of known and controlled volume. A high-sensitivity sensor measures pressures at the vertex of the cone. Using high-speed imaging we show how incoming liquid deforms the air bubble enclosed in such a geometry, and how an impact peak is registered inside the bubble{, which can be traced back to the impact of a liquid jet onto the pressure sensor}. We compare the measured pressures to a Bagnold model, and discuss the dominant resonances in the bubble. From visualisations of the deforming bubble, we also discuss the air-pocket's deformations, resulting from the presence of surrounding rigid geometry (such as corrugations in an LNG containment membrane).
\end{abstract}

\maketitle


\section{Introduction}

\begin{figure}
\includegraphics[width=0.5\linewidth]{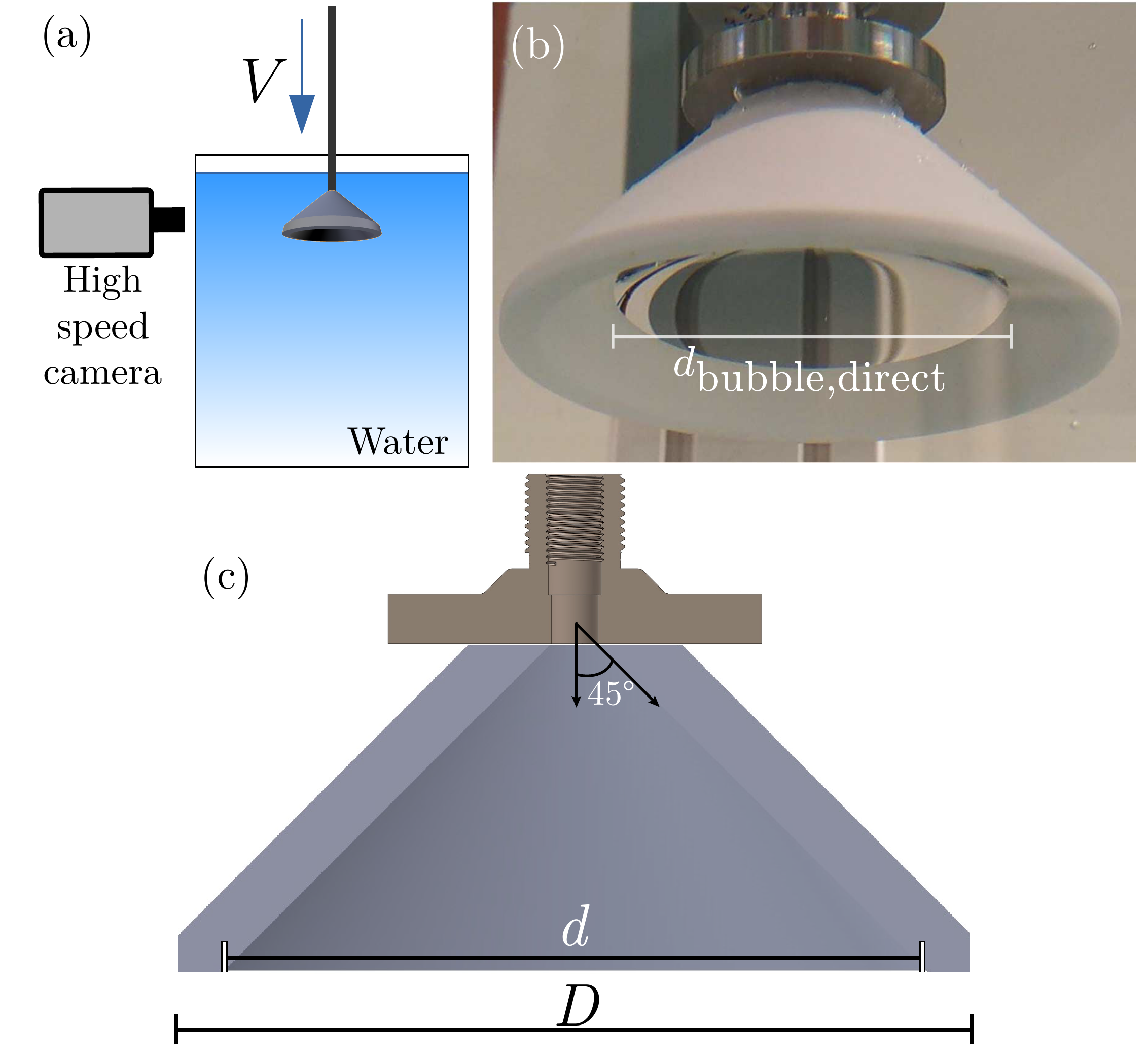}
\caption{Experimental setup: (a) {The inverted cone is placed in a water bath and moved downwards by means of a linear motor (not shown).} A high speed camera can be {used} to view the bubble's behaviour through a transparent cone. (b) Example of an air bubble trapped under the cone, {the surface of which} can be seen from the specular reflection {at the water-air interface}. (c)  {The pressure sensor is mounted in a} stainless steel disc {to which the 3D-printed cone is attached. Note that $D$ and $d$ represent the outer and inner diameter of the base of the cone, respectively. The sensor width is 5.55 mm.}} \label{setup}
\end{figure}

\begin{figure*}[h!]
\centering
\includegraphics[width=.99\linewidth]{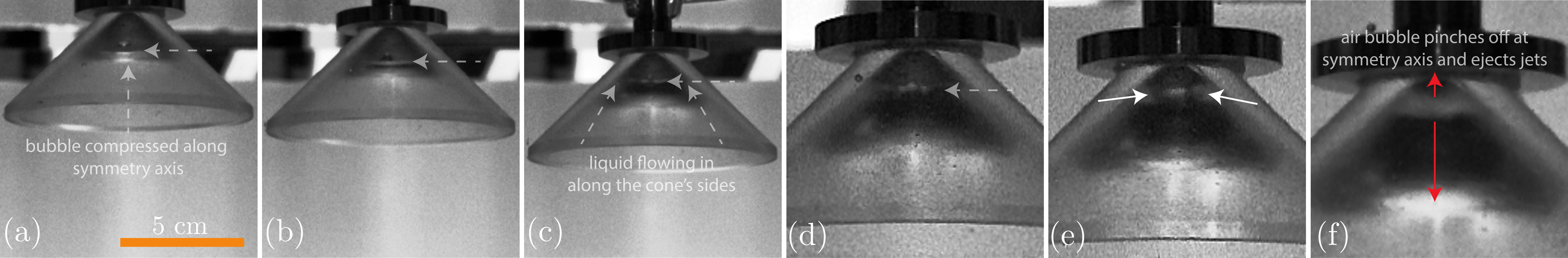}
\caption{Collapse of the bubble close to the vertex {of the cone}, when it is accelerated to move with $V = 1.7$ m/s. In this experiment, {each of the} panels shown above is separated by 7.17 ms {from the next.} The fluid {inside the cone} rises faster along the walls than {along} its symmetry axis. This results {in liquid flowing along the walls, accelerating and collapsing at the symmetry axis of the cone, possibly} at some {non-zero distance from the vertex} ({i.e., the} sensor). The horizontal and vertical dashed arrows show the bubble surface moving towards the cone vertex. {From the} pinch-off (panel (e)) {point a jet is ejected in the vertical direction.} Also see figure \ref{fig:pinchofffigs2} for another zoomed-in example.} \label{fig:pinchofffigs1}
\end{figure*}

In liquid-solid slamming events, air entrapment in between the former two phases is a common feature. It {may} occur due to air-cushioning, whereby the liquid surface deforms by a small amount due to high stagnation pressure, entrapping a thin air layer (\citet{verhagen1967, peters_2013,jainKH, jain_impulseJFM}). Alternatively it can occur when {either} the fluid or {the} solid phase has a concave surface, thereby fully entrapping a large pocket of air (\citet{mathai2015}). Examples of this type of entrapment can occur when an overturned wave slams onto a structure (\citet{bogaert2010, peregrine2003, kapsenberg, vanmeerkerk2020}), or during the re/-entry of water-skis with concave-shaped floats into water.\\

Air-pocket-mediated water slamming pressures have attracted attention from {the} sloshing and wave-slamming community due to the critical role that this mediating fluid can play in the final moments of hydrodynamic loading (\citet{denny1951, chan1988}). Indeed the first controlled study on wave-slamming onto walls by \citet{bagnold} concluded that the highest impact pressures were associated with the entrapment of a thin-air layer. This was also confirmed in experiments by \citet{Partenscky1988}, \citet{HATTORI199479} and \citet{richert1968}. While we now suspect this thin air-layer was a by-product of air-cushioning in front of a flat-faced wave (\citet{jain_impulseJFM}), the observations {inspired} the development of a model to understand the pressurisation of a gas pocket when it is compressed by an oncoming liquid column {known as the Bagnold model} (\citet{ancellin2012, brosset2013}). The entrapped air volume can have a variety of shapes, from thin and nearly-flat films (\citet{hicks_ermanyuk_gavrilov_purvis_2012, mayer_krechetnikov_2018, jainKH}), to thick {and compact} pockets (\citet{bogaert2010, kim_kim_kim_2021}) and thus {may} evolve in very different ways in the final stages of loading. While the detailed evolution of the entrapped air film generally does not {markedly} affect space-, or time-integrated quantities such as total force or the impulse on the target structure, the local pressures can be significantly influenced. An entrapped air pocket during slamming/loading is a bubble, and can have rich dynamics of its own when stimulated by the oncoming liquid mass. Indeed {vapor} bubbles may even cavitate and{, upon implosion,} produce even greater damage than mere added-mass of the liquid {may induce} (\citet{OKADA1995231, tagawa2018}). {In addition, } local pressures in the vicinity of an air-pocket are also sensitive to the local geometry of the solid target. An example of {such a} non-trivial target surface {shape} is found in the corrugations {of LNG containment systems, such as} MarkIII, by GTT (\citet{bogaert2010}). An air pocket entrapped against a corrugation would be {forcedly} squeezed into a tapering geometry where its deformation would exhibit more {complex} dynamics than {what would occur during the} compression of a spherical bubble in {the bulk of the liquid.} With {the aim of providing quantitative information of air pocket dynamics entrapped at a solid surface, }
in this paper we report experiments where an air pocket is contained inside an inverted cone immersed in water. The cone is subjected to an impact-like impulsive motion at a range of velocities. {We will discuss similarities and dissimilarities of our experiment in comparison with the idealised one-dimensional Bagnold model.} \\

We introduce the setup, describe the types of deformations the bubble exhibits, and analyse the pressures inside the bubble. We find that the presence of rigid boundaries around the {bubble} reduce its natural frequency of oscillation. The cone's tapering shape encourages an inward collapse of the bubble, resulting in {transient} variable peak pressures that are too large and too short-lived to be explained by Bagnold model.\\ 

The {overall pressurisation of the} bubble {that occurs once the cone is moving} at {a} constant velocity yields a situation {that is found to be} very close to Bagnold's model. {More specifically, the pressures occurring in this stage are  discussed in terms of the Bagnold number $S$.} As such, we use the model to estimate the {effective size of the} liquid column {that is accelerated during bubble compression, and compare it to the} experimental data. {Before describing the details of} the setup and methods in the next section{, we would like to stress that this} unique set of data could {only} be produced by ensuring that the air from the pocket did not leak from under the cone during the course of each measurement{, which placed strong restrictions on setup and measurements. Simply impacting a hollow object onto a water surface necessarily produces severe leakage of air from the pocket into the liquid (\citet{kimISOPE2017}) over a substantial period of time, which makes an analysis in terms of Bagnold's model extremely challenging.}\\

\section{Experimental setup}

The setup consists of a bath of water in a square tank of area $30 \times 30$ cm$^2$, and depth 60 cm. The liquid depth is kept constant at approximately 57 cm throughout experiments. A cone of opening angle 45$\degree$, {with an} internal diameter of 8.86 cm {at the base, and a} height of 4.16 cm, is inverted and submerged into the bath. This cone, the `impactor', has an opening at its vertex, where a stainless disc with flush-mounted Kistler 601CAA dynamic piezoelectric sensor is attached. The {disc and cone} are attached using Loctite SI595 silicone sealant which keeps the joint air-tight. The pressure sensor has a circular sensing area with a diaphragm {diameter} of 5.5 mm. All pressure measurements are done at an acquisition rate of 200 kHz.\\

The motion {of the impactor} is controlled with a NITEK linear actuator motor with an available stroke length of 700 mm. {After a brief acceleration phase, a} constant velocity was maintained over {a distance of} typically 35--45 cm. The initial submergence of the cone would entrap air {covering} its entire volume, {some of which} is expelled immediately when the cone is {moved}. Thus we control the amount of air volume trapped under the cone such that it does not accidentally leak during the course of different strokes. One way to achieve this is simply by repeatedly driving the assembly with a chosen velocity $V$, such that for a particular $V$, all `excess' air is thrown out from under the cone - the remainder volume of air then trapped remains constant during the course of experiments with the specific $V$. Velocities $V$ are varied between 0.1 and 2.9 m/s. {In all the $V$  covered in the present experiments, the acceleration phase spanned for approximately 3 ms.}\\

The air pocket is also observed from {the side using} a transparent cone, and a high speed camera at 10-16 kfps. Some examples are shown in figures \ref{fig:pinchofffigs1}, \ref{fig:pinchofffigs2} and {supplementary videos} (see Ancillary files in the arxiv submission).\\

\section{{Loading of the} air pocket and oscillations}

\begin{figure}[!]
\includegraphics[width=.55\linewidth]{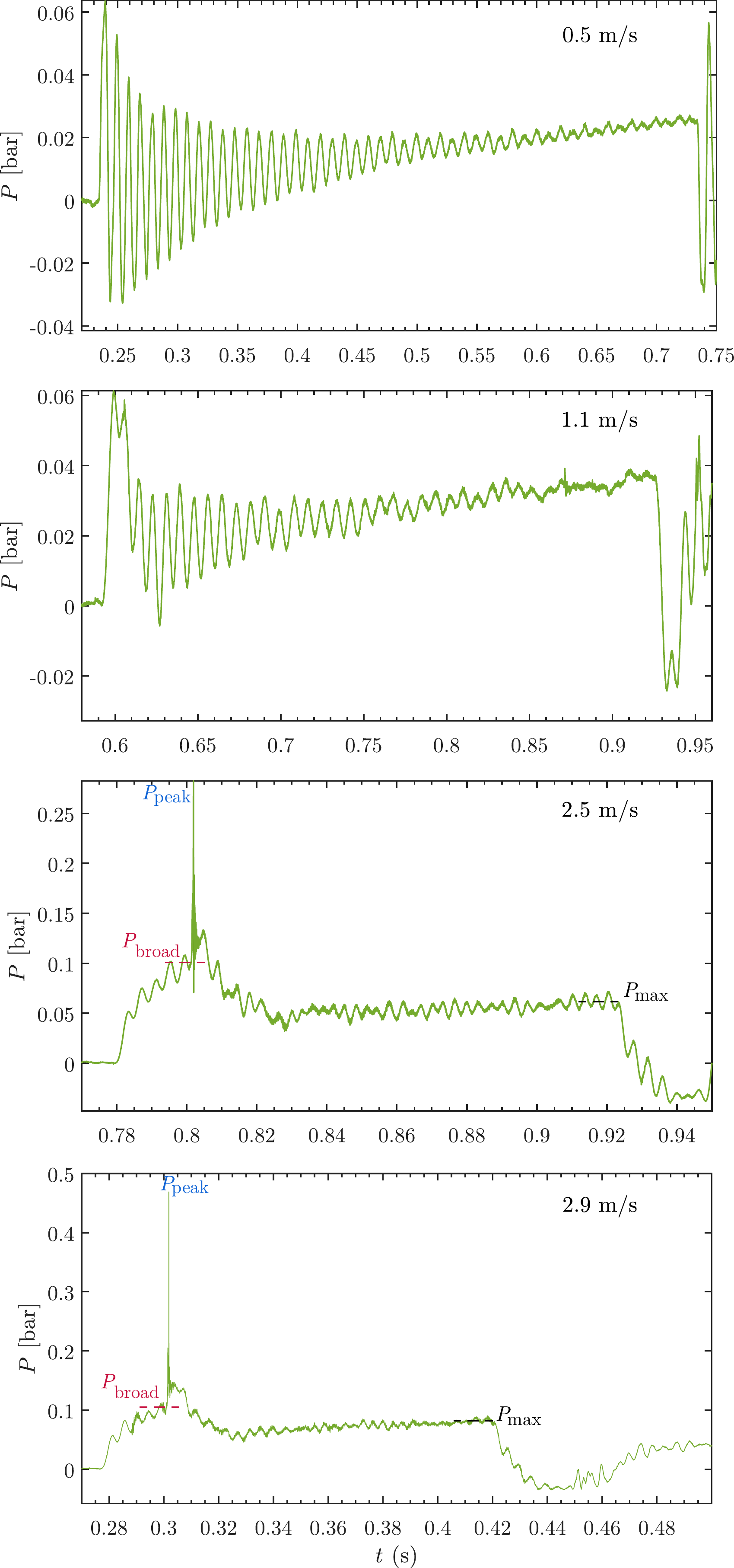}
\caption{{Time evolution of the pressure $P(t)$,} measured at the cone vertex in the presence of the trapped air bubble. The translucent blue signals are the raw data, and the green signals are the low-pass filtered data at 1/8 of the sampling frequency. The three panels are shown for experiments at different velocities $V =  0.5$, $1.1$ and $2.5$ m/s {and for different bubble sizes.} {The horizontal red dashed lines indicate the value of the broad peak pressure $P_\text{broad}$, the peak pressure due to collapse of the bubble is denoted by $P_{\text{peak}}$, and the stagnation pressure (denoted $P_{\text{max}}$) in the bubble is shown by the black dashed line.}}\label{fig:pressuretimeseries}
\end{figure}

An example of the evolution of the air-pocket's surface is shown in figure \ref{fig:pinchofffigs1}. Typically a high stagnation pressure in the liquid would compress the bubble along the symmetry axis. This compression at {the} center results in the air mass being squeezed out along the cone's sides ({as can be observed in} panels a--c in figure \ref{fig:pinchofffigs1}). {This outward motion of the air phase is counteracted by liquid entering the cone, which is being forced to converge towards the symmetry axis by the shape of the cone (panels d--f). This results in} a narrowing of the air-pocket {close to or at} the disc. This narrow gap {may} pinch off and {emit liquid} jets (\citet{gekle2009jet, bergmann2006}) {along the symmetry axis, the impact of which creates} a large {and sharp} pressure peak (henceforth called $P_{\text{peak}}$) on the sensor (see the sharp pressure peak in the time series in last panel of figure \ref{fig:pressuretimeseries} at 0.8 s). {We will now discuss the features of the pressure signal in greater detail.}\\

Typical pressure time series at $V =  0.5$, $1.1$ and $2.5$ m/s are shown in figure \ref{fig:pressuretimeseries}. The unfiltered signals are shown in blue with partial opacity {and the} green time series are low-pass filtered at a cut-off frequency of 1/8th of the acquisition frequency. The initial increase of pressure results from a combination of the assembly's acceleration and the bubble's compression. The time series {at the highest velocity ($V=2.5$ m/s)} shows that the filtering still retains the shortest time scale ($\sim \mathcal{O}(10^{-4})$s) signals (the sharp peak at {$t = 0.8$ s}) {that correspond to the pinch off of the bubble and the creation of the jet.} The next stage is {reached} when the cone moves with a constant velocity, and the recorded pressure evolves to oscillate about {an almost} steady value. The final stage shows a drop in the pressure (in figure \ref{fig:pressuretimeseries}, at $0.72$ s for $V = 0.5$ m/s, $0.92$ {s} for $V= 1.1$ m/s and approximately $0.92$ {s} for $V = 2.5$ m/s), which corresponds to the cone-disc assembly {decelerating and} coming to rest. Notice that, starting with the assembly's initial acceleration, the pressure {oscillates} with a specific frequency {which persists} even after the assembly ceases to move.\\

\begin{figure}
    \includegraphics[width=.62\linewidth]{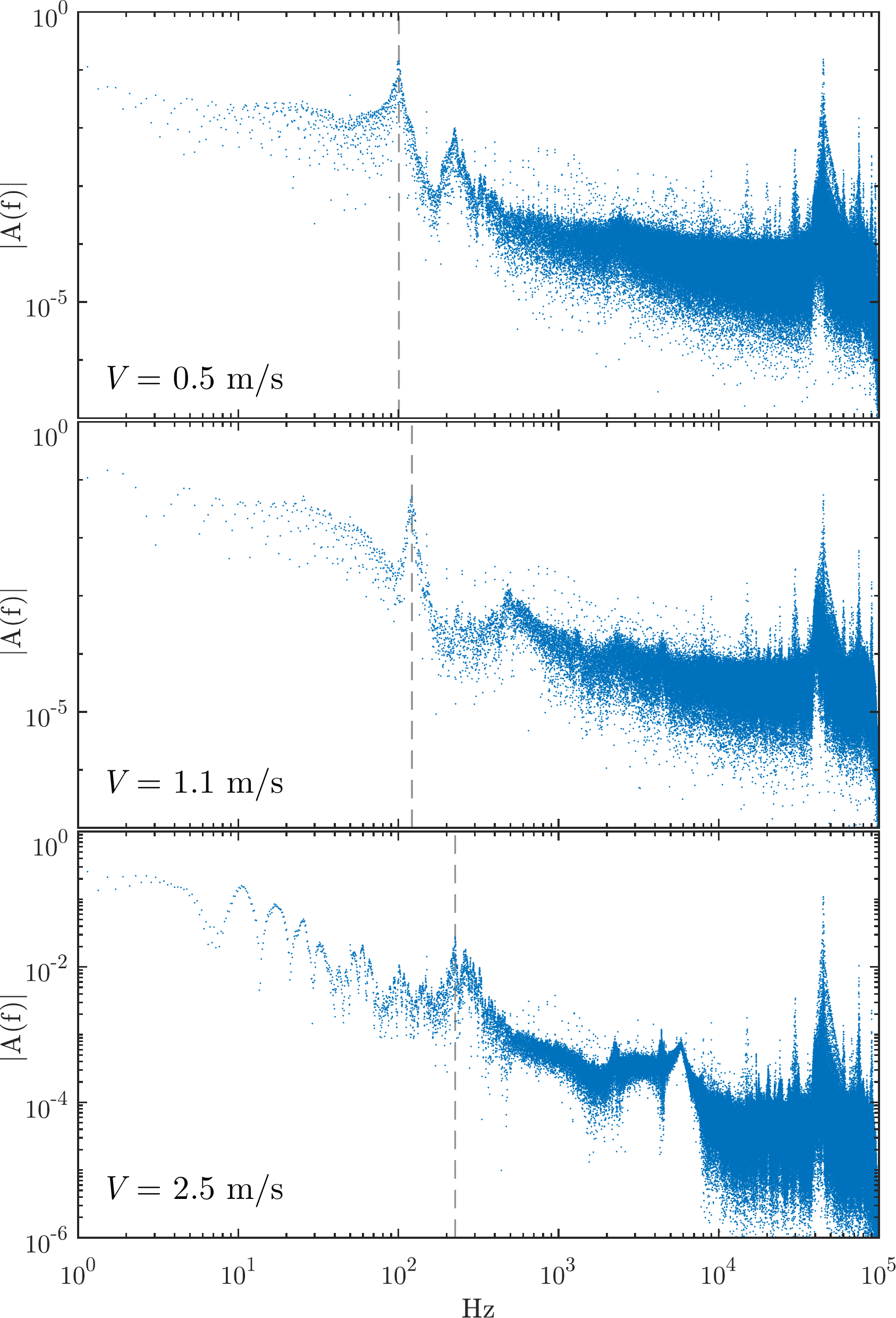}
    \caption{One-sided amplitude spectra obtained from Fourier transforms of the pressure time series {of figure  \ref{fig:pressuretimeseries}}. {The d}ashed vertical line in each panel shows how the Minnaert resonance peak was identified to obtain the effective bubble radius from its oscillation frequency, when subjected to motion.}    \label{fig:FFTs}
\end{figure}

{In figure \ref{fig:FFTs} the} amplitude spectra obtained from fast Fourier transforms of the pressure time series {of figure \ref{fig:pressuretimeseries}} are shown. From such spectra, we deduce the natural frequency of the {oscillating} bubble. Although approximately 60\% of this conical bubble's surface area is in contact with the rigid cone{, the result of} \citet{minnaert1933} for the natural frequency $f$ of {an oscillating} bubble may still be used (\citet{blue1967}) despite some lowering of the natural frequency (\citet{payne2005}). From \citet{minnaert1933},
\begin{equation}
    f = \frac{1}{2 \pi R } \left( \frac{3 \gamma {p_0}}{\rho_w}\right)^{1/2},
\end{equation}
where $R$ is the radius of the bubble, $\gamma$ is the polytropic coefficient, {$p_{0}$} is the ambient pressure, and $\rho_w$ is the density of water. For {an} air {bubble in water}, $\gamma = 1.4$, {and} the above relation simplifies to $R_{\text{Minnaert}} = 3.26/f$ m.\\

\begin{figure}
\includegraphics[width=.7\linewidth]{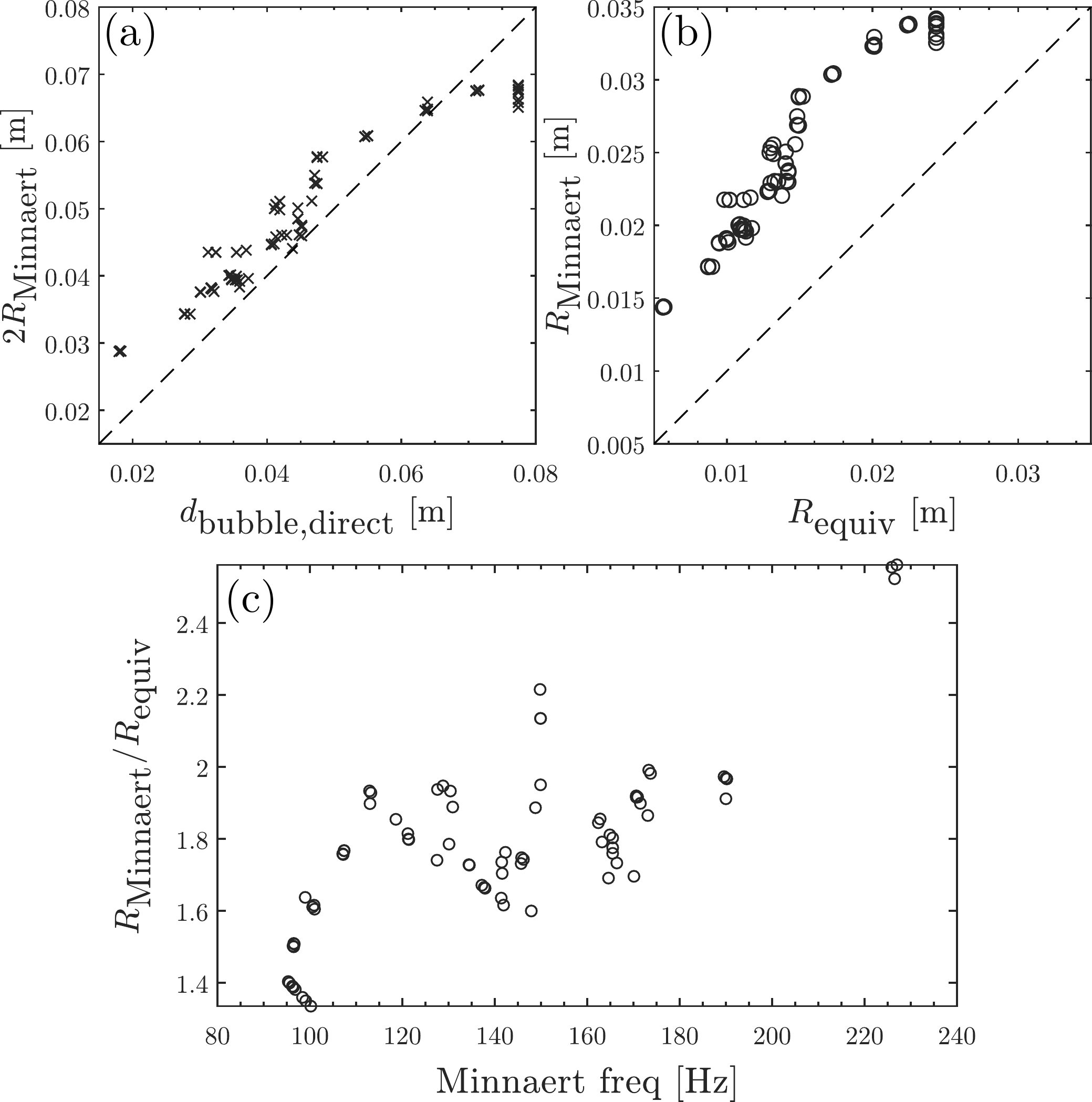}
\caption{(a) Comparison of {the diameter $d_{\text{bubble,direct}}$ of the base of the conical bubble,} measured directly (visually) {in the setup (figure \ref{setup}(d))} and the {acoustic} bubble {diameter $2R_{\text{Minnaert}}$,} estimated from its Minnaert resonance frequency. (b) Comparison of {the acoustic} bubble radius {$R_{\text{Minnaert}}$,} estimated from {its} Minnaert resonance frequency to the effective bubble radius {$R_{\text{equiv}} = (3U/4\pi)^{1/3}$ estimated from the conical volume $U$} of trapped air. (c) The {acoustic} method overestimates the bubble size: the extent of this is shown in comparison to $R_{\text{equiv}}$, and its variation with the measured bubble frequencies $f$ is shown.}
\label{fig:directbubsize_fftsize}
\end{figure}

The {acoustic bubble diameters} thus deduced from Fourier spectra ($2R_{\text{Minnaert}}$) are compared to the {diameter $d_{\text{bubble,direct}}$ of the base of the} conical bubble, measured {directly} from the setup {by visual inspection} (figure \ref{fig:directbubsize_fftsize}, panel (a)). {Alternatively,} $R_{\text{Minnaert}}$ {is compared} with the {effective radius obtained from the volume $U$ of the conical bubble using $R_{\text{equiv}} = (3U/4\pi)^{1/3}$} in panel (b). This result is striking for two reasons: firstly in panel (a), the excellent agreement of $2R_{\text{Minnaert}}$ and $d_{\text{bubble,direct}}$, where the latter is a very rough indicator of the bubble's oscillating surface, and not the correct representative of the bubble's volume. Secondly in panel (b), $R_{\text{Minnaert}}$ {is} found to be consistently {larger} than $R_{\text{equiv}}$, indicating that the bubble is oscillating at a lower frequency than it would have done in the absence of the rigid boundaries. The disagreement in panel (b) is then an indicator of the damping of the {bubble} oscillations. Such damping can be either viscous or thermal {in origin, which both} would increase with the frequency. We try to {quantify} this in panel (c) of figure \ref{fig:directbubsize_fftsize}, by plotting the {ratio $R_{\text{Minnaert}}/R_{\text{equiv}}$} against bubble frequency. The trend in panel (c) indicates that the damping indeed {roughly} increases with the bubble frequency.

\section{{Air pocket pressure at constant velocity}}

\begin{figure}
    \centering
    \includegraphics[width=.6\linewidth]{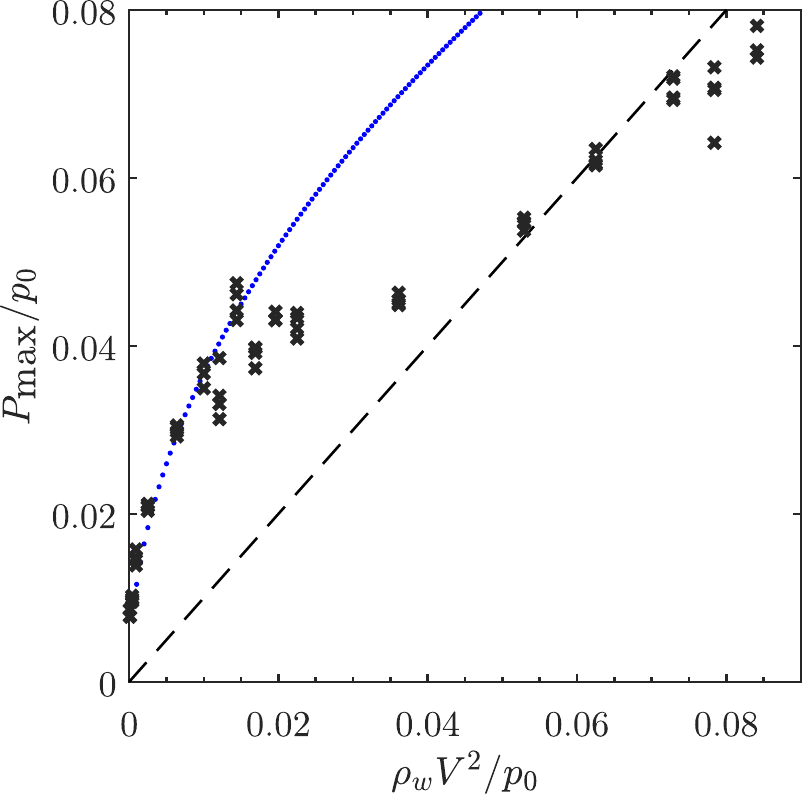}
    \caption{{The plateau value of the measured pressure $P_\text{max}$, non-dimensionalised with the atmospheric pressure $p_0$ is plotted as a function of the stagnation pressure $\rho_w V^2$, also non-dimensionalised with $p_0$ (black symbols). On the dashed line $P_\text{max}$ and $\rho_w V^2$ are equal. The blue dotted line is a square root fit of the data for small impact velocities} {of the form $P_\text{max}/p_0 = a \sqrt{\rho_w V^2/p_0}$  (see equations \eqref{pmaxapprox} and \eqref{eqn:pmaxsquareroot}). Here, the fitting range has been restricted to data with $\rho_w V^2 \leq 0.15 p_0$, leading to $a = 0.37$.}     }
    \label{fig:stagnation}
\end{figure}

{Despite the presence of the oscillations,} the pressurisation {of the bubble is readily measured}  from the time series. For a given {velocity} $V$, {the bubble is under maximum compression when the cone} nears the end of its prescribed stroke. We {measure this} steady pressure value about which {the oscillation takes place} prior to deceleration, and {call it} $P_{\text{max}}$.  {Naively, one may expect this pressure to be proportional to the stagnation pressure of the bubble moving through the liquid, i.e., $P_\text{max} \approx \rho_w V^2$. In figure \ref{fig:stagnation} we therefore plot  $P_\text{max}/p_0$ as a function of $\rho_w V^2/p_0$. Whereas for larger impact velocities $V$ the data lie close to the line $P_\text{max} = \rho_w V^2$, the measured pressures are observed to be much larger than the stagnation pressure for small impact speeds. We will now investigate to what extent this observation is connected to the dynamic pressurization of the gas pocket as described by the Bagnold model. To do so, we first} briefly revisit the {one-dimensional} piston model of \citet{bagnold} {in the next subsection}.\\

\subsection{Brief {summary} of 1D piston model}

A piston with one end open {to the environment} contains a liquid 
column of length $K$, which moves to compress a gas column against the closed end of the tube. The gas column has initial length of $x(t=0) = x_0$, and is initially at {the atmospheric} pressure $p_0$. The liquid column moves {with an initial velocity {$dx/dt = -V$ at time $t = 0$}} to adiabatically compress the gas column{, during which the} 
pressure in the gas follows the adiabatic 
law $P_{\text{gas}}(t) = p_0 \left( x_0/x(t) \right)^\gamma$. 
The resulting equation of motion of the 
{liquid column is}
\begin{equation}
   { \rho_w K \frac{d^2 x}{dt^2} = -p_0 + P_{\text{gas}} = -p_0 \left( 1 - \left( \frac{x_0}{x} \right)^\gamma \right)\,,}
\end{equation}
subject to initial conditions $x(0)=x_0$ and {$dx/dt = -V$ at $t = 0$}. We non-dimensionalise $x$ and $t$ as \begin{equation}
    \tilde{x} = \frac{x}{x_0} \text{ and } \Tilde{t} = \frac{V t }{x_0}.
\end{equation}
The nondimensionalised equation of motion of the boundary is then \begin{equation}
    \frac{d^2 \Tilde{x}}{d \Tilde{t}^2} = - \frac{p_0 x_0}{\rho_w V^2 K} \left( 1 - \tilde{x}^{-\gamma}\right),
\end{equation}
where the prefactor on the right hand side can be identified as Bagnold number $S$ mentioned in the previous section. {The energy integral} is found by multiplying the above expression with $d\tilde{x} = \frac{d\tilde{x}}{d\tilde{t}} d\tilde{t}$ and integrating once. Bringing all the terms to one side, 
\begin{equation}\label{energyeq}
    dE \equiv d \left( \frac{1}{2} \left( \frac{d \tilde{x}}{d \tilde{t}} \right)^2 + S\tilde{x} \left( 1 + \frac{1}{\gamma-1} \tilde{x}^{-\gamma} \right) \right) =  0,
\end{equation}
implying that {the total} energy $E$ is a constant. At $\tilde{t}=0$, \begin{equation}\label{energyinitial}
  E_{\tilde{t}=0} = 1/2 + \gamma S/(\gamma - 1).  
\end{equation}  
At the instant of maximum compression $\tilde{t} = \tilde{t}_{\text{max}}$, $\tilde{x}(\tilde{t}_{\text{max}}) \equiv \tilde{x}_{\text{min}}$, and $\frac{d\tilde{x}}{d \tilde{t}}(\tilde{t}_{\text{max}})=0$, such that the energy {equals} 
\begin{equation}\label{energyfinal}
    E_{\tilde{t}=\tilde{t}_{\text{max}}} = S \tilde{x}_{\text{min}} \left( 1 + \frac{1}{\gamma -1} \tilde{x}_{\text{min}}^{-\gamma} \right). 
\end{equation}

\begin{figure}
    \centering
    \includegraphics[width=.7\linewidth]{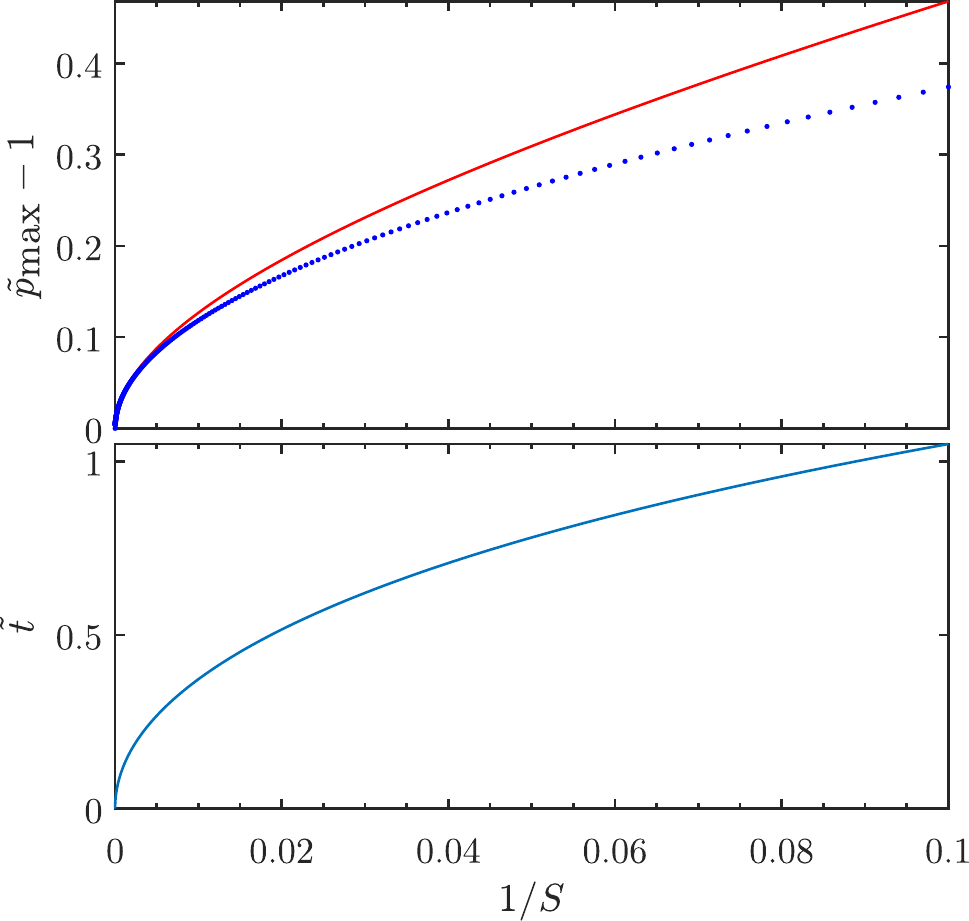}
    \caption{{(a)} Solution to equation \eqref{bagnoldnondimrelation}{, representing the theoretically predicted maximum dimensionless pressure in the air pocket $\tilde{p}_\text{max}-1$ as a function of the inverse Bagnold parameter, i.e., $1/S$. The blue dotted curve represents the approximate solution \eqref{pmaxapprox} for large $S$ (i.e., $1/S \ll 1$). (b) Dimensionless time $\tilde{t}_\text{max}$ at which maximum compression of the air pocket is reached, also versus $1/S$.} 
    }     \label{fig:bagnoldnondim}
\end{figure}

{Due to e}nergy conservation \eqref{energyinitial} {equals} \eqref{energyfinal}, {and defining $\tilde{p} \equiv P_{\text{gas}}/p_0$, the adiabatic compression law can be written as  $\tilde{p} =  \tilde{x}^{-\gamma}$. Since the maximum pressure $\tilde{p}_{\text{max}}$ occurs at $\tilde{t}_{\text{max}}$, we find that $\tilde{p}_{\text{max}} =  \tilde{x}_{\text{min}}^{-\gamma}$,} resulting in the following expression relating the maximum pressure inside the bubble to $S$: \begin{equation}\label{bagnoldnondimrelation}
    \tilde{p}_{\text{max}}^{-1/\gamma} \left( 1 + \frac{1}{\gamma -1} \tilde{p}_{\text{max}}\right) = \frac{1}{2S} \left( 1 + \frac{2 \gamma}{\gamma -1} S \right).
\end{equation}

\subsection{{Large $S$ regime}}

All the experiments covered here, {are done at a small impact velocity, i.e., they satisfy 
$\rho_w V^2/p_0  
< 0.08$ and are in the regime where $S$ is large ($S \gg 1$). This implies that 
$\tilde{p}_{\text{max}}$ is expected to be very close to $1$, such that one may write $\tilde{p}_{\text{max}} = 1 + \varepsilon$ with $\varepsilon \ll 1$. Inserting this expression in \eqref{bagnoldnondimrelation} and expanding the left hand side up to second order in $\varepsilon$ leads to
\begin{equation}
   \frac{1}{\gamma-1} \left(\gamma + \frac{\gamma-1}{2\gamma}\varepsilon^2 + O(\varepsilon^3)\right) = \frac{1}{2S} \left( 1 + \frac{2 \gamma}{\gamma -1} S \right)\,,
\end{equation}
which can be solved straightforwardly as
 \begin{equation}\label{pmaxapprox}
   \tilde{p}_{\text{max}} - 1 = \varepsilon \approx \sqrt{\frac{\gamma}{S}}\,.
\end{equation}}

{In addition, one may solve the energy equation \eqref{energyeq} and \eqref{energyinitial}
\begin{equation}
 \frac{1}{2} \left( \frac{d \tilde{x}}{d \tilde{t}} \right)^2 + S\tilde{x} \left( 1 + \frac{1}{\gamma-1} \tilde{x}^{-\gamma} \right) = \frac{1}{2} + \frac{\gamma}{\gamma - 1}S
\end{equation} 
for $d\tilde{x}/d\tilde{t}$, from which we can write the time needed to reach maximum compression as an integral
\begin{eqnarray}
\tilde{t}_\text{max} &=& \int_{\tilde{x}=\tilde{x}_\text{min}}^1 \frac{d\tilde{x}}{\sqrt{1 + \frac{2\gamma}{\gamma-1}S -2S\tilde{x}(1+\frac{1}{\gamma-1}\tilde{x}^{-\gamma}) }}\\
&=& \int_{\tilde{p}=1}^{\tilde{p}_\text{max}} \frac{d\tilde{p}}{\gamma\tilde{p}^{(\gamma+1)/\gamma}\sqrt{1 + \frac{2\gamma}{\gamma-1}S -2S\tilde{p}^{-1/\gamma}(1+\frac{1}{\gamma-1}\tilde{p}) }}\,.
\end{eqnarray} }

{Solutions to equation \eqref{bagnoldnondimrelation} are shown in figure \ref{fig:bagnoldnondim}a as a function of $1/S$ together with the approximation of \eqref{pmaxapprox}. Clearly, in the region of interest ($\tilde{p}_\text{max}-1<0.1$) the approximation is quite accurate such that we fit the smaller impact speed values plotted in figure \ref{fig:stagnation} to a square root, rewriting \eqref{pmaxapprox} as}
\begin{equation}\label{eqn:pmaxsquareroot}
{\frac{p_\text{max}}{p_0} \approx \sqrt{\frac{\gamma K}{x_0}}\sqrt{\frac{\rho_w V^2}{p_0}}\,.}
\end{equation}
{From this fit (see figure \ref{fig:stagnation}) we find that $a = \sqrt{\gamma K/x_0} \approx 0.37$,} {and using that $\gamma = 1.4$ for air, we can estimate the ratio of the vertical extent of accelerated liquid $K$ and the effective bubble size $x_0$} {as $K/x_0 = a^2 / \gamma \approx 0.096$. We will discuss the significance of this result in the next subsection.}\\

\begin{figure}
    \centering
    \includegraphics[width=.8\linewidth]{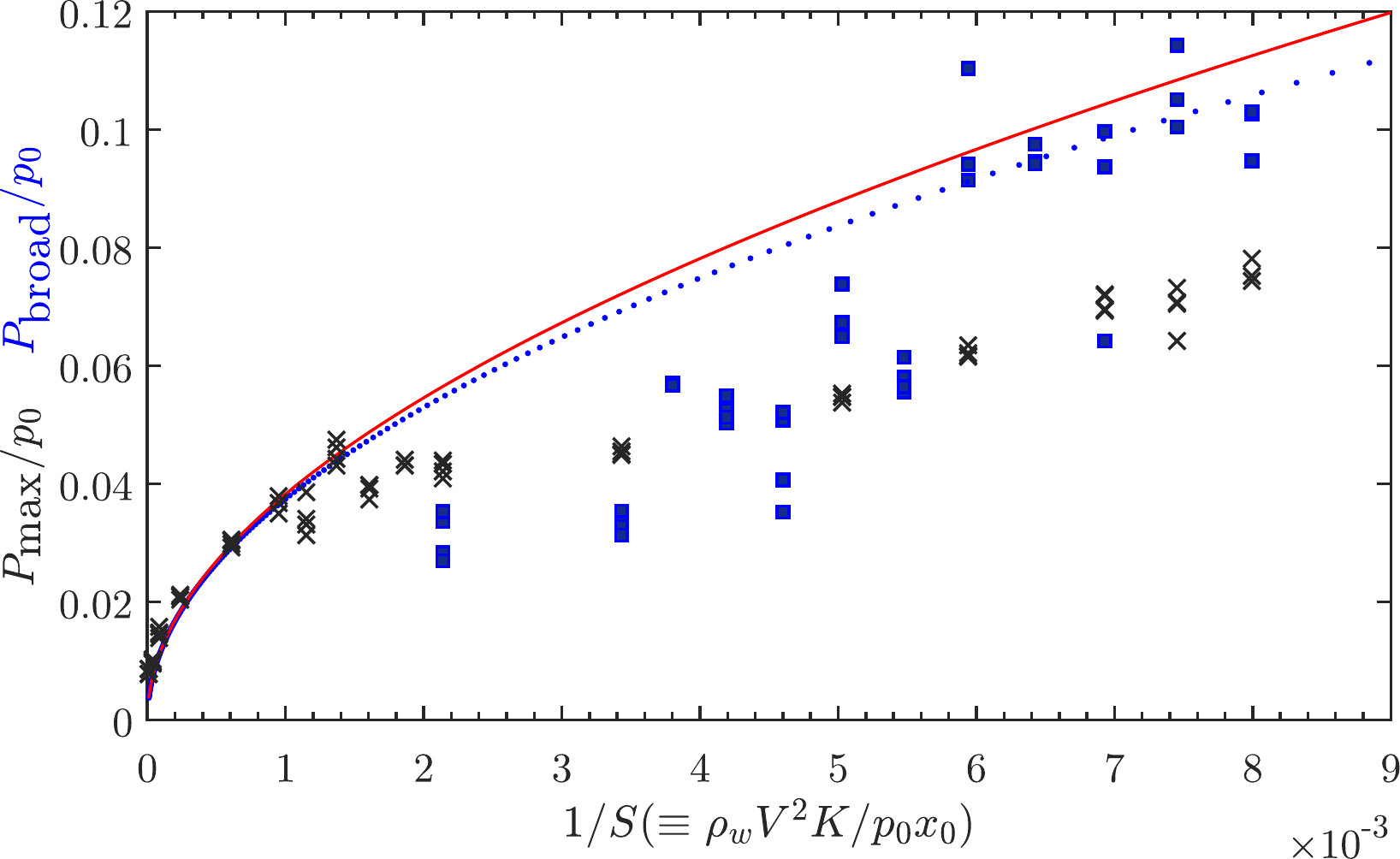}
    \caption{{The plateau value of the measured pressure $P_\text{max}$ (black crosses) and the broad peak value $P_\text{broad}$ (solid blue squares), both non-dimensionalised with the atmospheric pressure $p_0$, plotted versus the inverse Bagnold number $1/S$, using $K/x_0$ from the fit obtained in figure \ref{fig:stagnation}. Here, $P_{broad}$ is determined as the bubble pressure close to the time when the sharp peak is produced (as visible in the last panel of figure \ref{fig:pressuretimeseries} at $t = 0.801$ s). The black solid curve represents the solution of the full Bagnold model \eqref{bagnoldnondimrelation} and the blue dotted line the square root approximation of \eqref{eqn:pmaxsquareroot}.} 
  }
    \label{fig:steadypressureinbubble}
\end{figure}

{In figure \ref{fig:bagnoldnondim}b, the dimensionless time needed to reach full compression of the air pocket is plotted against the inverse Bagnold parameter $1/S$. For the region of interest, $\tilde{t}_\text{max} \approx 0.1-0.3$, which needs to be multiplied with $x_0/V$ to obtain dimensional times, which for the experiments presented here range between $1$ and $15$ ms.}\\

\subsection{{Comparison with experiments and discussion}}

{In figure \ref{fig:steadypressureinbubble} we plot the same data as was plotted in figure \ref{fig:stagnation} (black crosses), and added the initial broad peak pressure values $P_\text{broad}$ for all cases where we could obtain it from the time series (solid blue symbols, cf. figure \ref{fig:pressuretimeseries}). Here, $P_\text{broad}$ is the value of the broad first peak, taking into account that there may be a second, sharp peak, caused by the impact of a liquid jet onto the sensor (as in the lower panel in figure \ref{fig:pressuretimeseries}) which is discarded when determining $P_\text{broad}$. On the horizontal axis we used the inverse Bagnold number $1/S$, where the quantity $K/x_0 \approx 0.096$ was taken from the square root fit in figure \ref{fig:stagnation}. The blue dotted line represents the square root fit \eqref{pmaxapprox} and the solid red line the full solution of equation \eqref{bagnoldnondimrelation}.}\\

{Remarkably, the $P_\text{broad}$ data lie very close to that predicted by the Bagnold model for large values of the velocity $V$, which indicates that for these cases $P_\text{broad}$ indeed represents the maximum pressure predicted by the Bagnold model, and that the experimental setup used in these experiments can at least in principle be used to test the response of a gas pocket against such a simple one-dimensional model. However, upon inspecting the data more closely, some features ask for a more in depth discussion.}\\

{The first feature that deserves our attention is the small value of $K/x_0$, namely approximately equal to $0.096$. This implies that, identifying $x_0$ with some measure of the bubble size, e.g., $R_{\text{Minnaert}}$, we find values for $K$ to lie in the range of $1.4$--$3.3$ mm. This value is surprisingly small, when realising that $K$ is expected to indicate the size of the added mass of liquid, which for the inverted cone could be estimated as the sum of the liquid present within the hollow cone and a hemispherical mass of liquid attached to the base of the cone. This would lead to a value of $K$ that is at least an order of magnitude larger than the ones estimated from applying Bagnold's model.}\\

{Connected to the above we may ask ourselves the question in how far the one-dimensional Bagnold model is applicable to our experiment. The first difference is that the ullage pressure, which in the model was taken to be equal to the atmospheric pressure $p_0$ is in fact equal to the stagnation pressure for a body moving at a velocity $V$ through the liquid, i.e., $p_0 + \rho_w V^2$. The Bagnold model is in fact easily modified to include this slightly higher ullage pressure by replacing $p_0$ by $p_0 + \rho_w V^2$, and we only have not included it to keep the model as transparent as possible.}\\

{The second difference however has much larger consequences, and that is that we need a short but finite time to accelerate our cone to its final velocity $V$. To evaluate the impact of the acceleration phase, we can imagine bringing the cone very slowly to its final velocity $V$. In that case the pressure inside the gas pocket would also very slowly rise from atmospheric to its stagnation value, without  any sign of the compression predicted by the Bagnold model. So it stands to reason that the acceleration phase has a significant effect on the maximum pressure that is reached in the air pocket, and that it should be kept as short as possible to closely approach the conditions of the Bagnold model. It is in fact plausible that the small $K/x_0$ value found in the current experiments are connected to a smaller compression due to the finite time that the linear motor needs to accelerate up to the target velocity $V$.}\\

{These observations lead to a proposal for an improved experimental design that will be discussed in the conclusions section. But first we will turn to discussing the pressure spike that is caused by the impact of a liquid jet onto the sensor surface.}

\section{Bubble collapse at higher velocities}

\begin{figure}
    \centering
    \includegraphics[width=.65\linewidth]{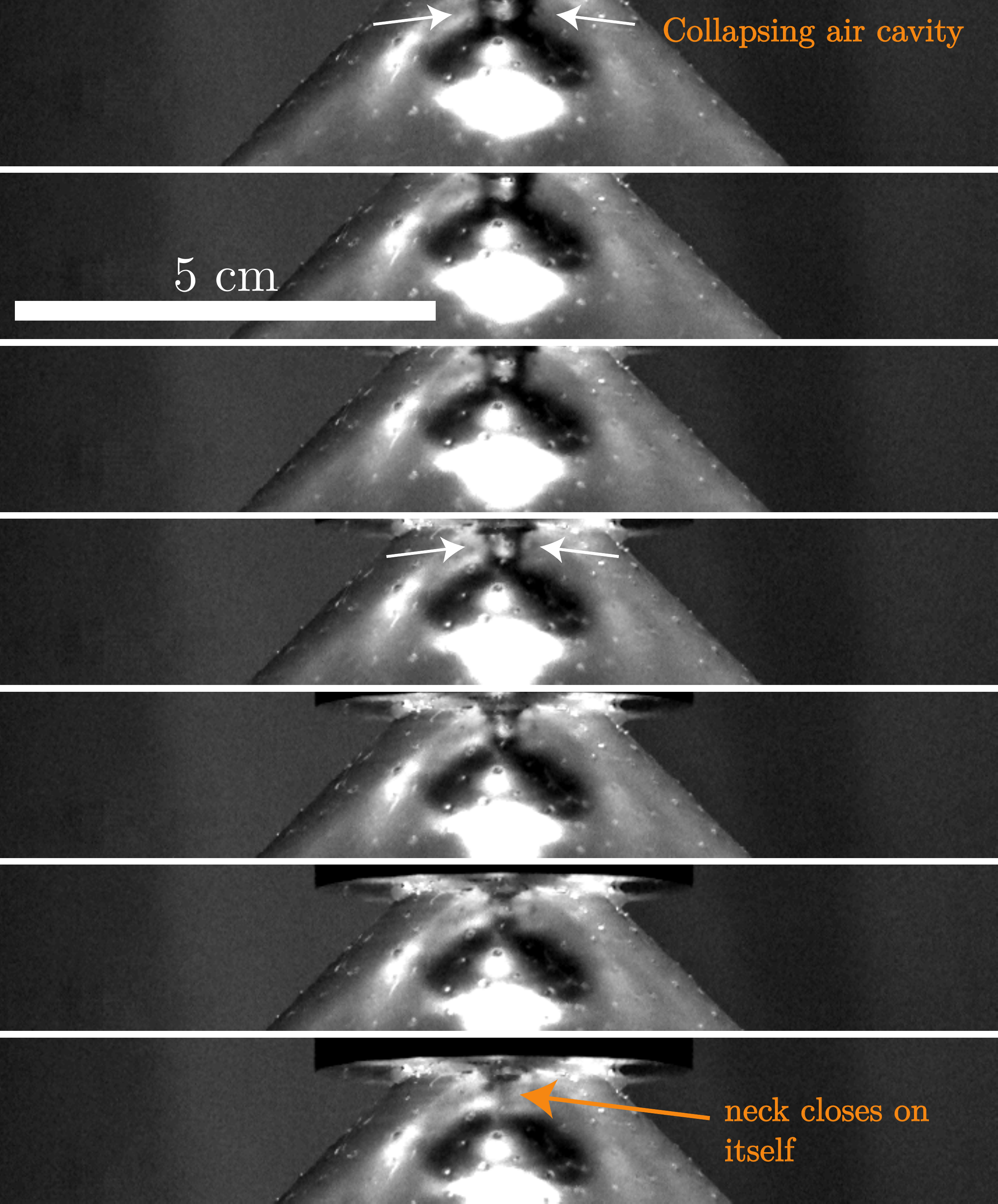}
    \caption{Collapse of the bubble close to the {vertex of the cone,} when it is accelerated to move with $V = 1.75$ m/s. In this experiment, the panels shown above are each separated by 0.625 ms. {Due to the geometry of the cone, the} fluid rises faster along {its} 
    walls than the bubble {is compressed} along its symmetry axis. This results in the bubble pinching-off at {or close to} 
    the vertex ({i.e., at the pressure} sensor). The pinch-off ejects two vertical jets in opposite directions {along the symmetry axis, which produces a short pressure spike in the sensor.}}
    \label{fig:pinchofffigs2}
\end{figure}

The above {data} analysis and {interpretation} focused only on {that part of the} impact pressure {signal that could be attributed to a compression of the} air pocket {as a whole}. However, situations {may occur} where the {liquid flow around the} 
air pocket {gives rise to singular events that show up as a short and sudden pressure increase in the pressure signal.} {More specifically, we} observed in the present experiments that the tapering geometry enclosing the bubble made {the liquid} pinch off at {or close to} the vertex {of the cone, an} example {of which} is shown in figure \ref{fig:pinchofffigs2}. {Here, one can observe how the top part of} the air bubble {is constricted by the radial inflow of liquid that eventually leads to a} pinch-off event {in which} jets {are ejected} from the pinch-off location {in the vertical direction.} We noticed that sharp pressure peaks such as that shown in figure \ref{fig:pressuretimeseries} for $V = 2.5$ m/s always occurred immediately after the {moment at which} such a pinch-off was seen from the side view. {Experimental videos {can be accessed from Ancillary files in the arxiv submission}}.\\ 

\begin{figure}
    \centering
    \includegraphics[width=.6\linewidth]{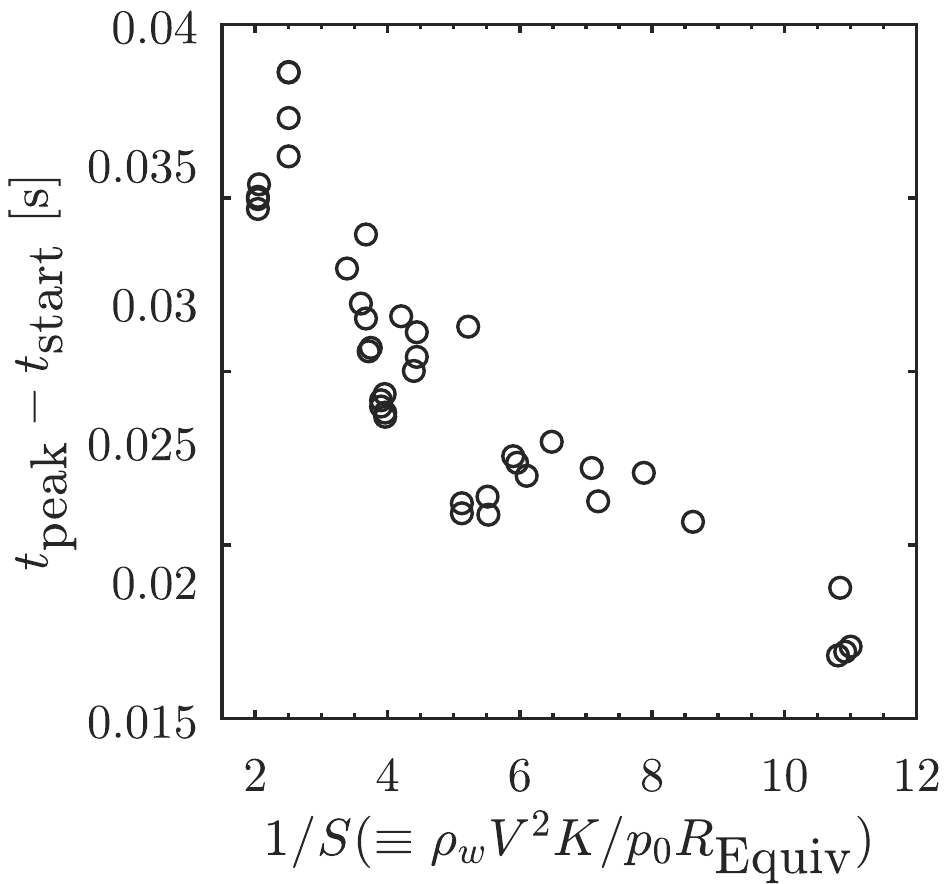}
    \caption{{The time interval $t_\text{peak} - t_\text{start}$ between} the {starting time $t_\text{start}$) of the motion of the} cone {and the time $t_\text{peak}$ at which the sharp pressure peak connected to the pinch-off and jet-formation process} is recorded{, is} plotted {versus the} {inverse Bagnold number $1/S$.} }     \label{fig:pinchofftimes}
\end{figure}

The pressurisation of the bubble starts from the moment the cone begins accelerating. We measured the time {difference} 
from the start of {the motion of the cone} 
($t_{\text{start}}$) until {the moment that} 
the sharp pressure peak $P_{\text{peak}}$ from the bubble's pinch-off appears (denoted as $t_{\text{peak}}$). The time {interval} 
$t_{\text{peak}} - t_{\text{start}}$ {is} 
plotted as a function of $\rho_w V^2$ and $1/S$ in figure \ref{fig:pinchofftimes}. {Note that in all cases the initial acceleration phase spans less than 15\% of the measured time difference, such that the dynamics of the pinch-off process is dominated by the liquid flow at a constant velocity $V$.} It is clear from figure \ref{fig:pinchofftimes}(a) that a higher velocity $V$ results in a more rapid collapse of the bubble. However, on including the influence of bubble size $R_{Equiv}$ in panel (b), we see the spread {in the} data {becomes less and it lies closer to} a single curve.\\ 

\begin{figure}
    \centering
    \includegraphics[width=.5\linewidth]{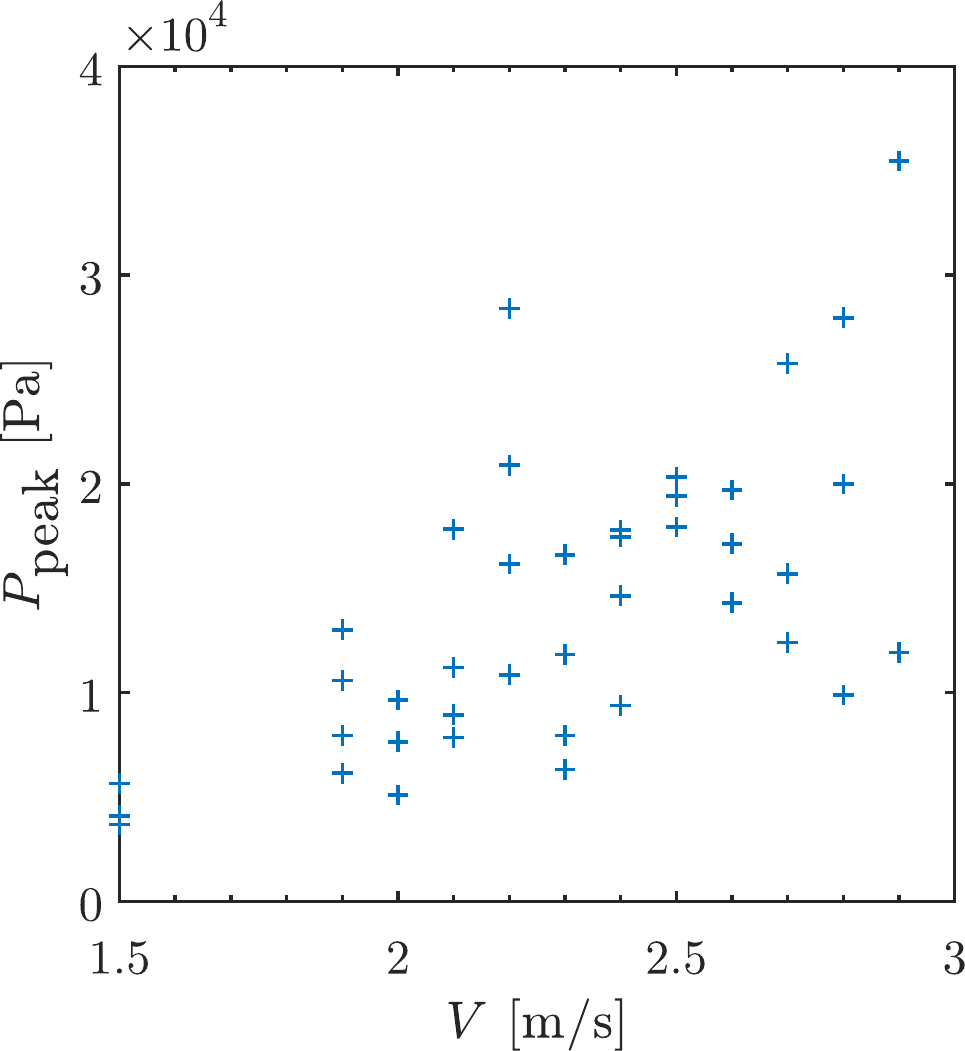}
    \caption{The {magnitude of the} impact {pressure} peaks $P_\text{peak}$ resulting from the 
    pinch-off {of the bubble} ({cf.}  
    panel (c) of figure \ref{fig:pressuretimeseries}) are plotted against {the impact velocity} $V$. The {large} 
    spread in the data shows the inherent variability {of the pinch-off and jet-formation processes.}} 
    \label{fig:variabilityofimpactpeaks}
\end{figure}

Finally, we look at the {magnitude of the} peak pressures created by the 
{pinch-off event} described above. The {pressure} peaks have 
temporal widths of the order of $10^{-4}$ s, which is similar to those found in impacts of a flat disc on water (\citet{jain_impulseJFM}). These {short durations} 
are {in} clear contrast {with} the impact pressures discussed above, that were mediated by the {increased gas pressure inside the} bubble {as a whole}. We plot the {magnitude of the pressure} peaks against {impact velocity} $V$ in figure \ref{fig:variabilityofimpactpeaks}. The data shows a large spread despite being well resolved in time. We ascribe the spread to variability {in the liquid flow that leads to the pinch-off event, leading to, e.g., changes in the degree of axisymmetry and the precise location of the collapse.}

\section{Conclusions}

We have shown how an air bubble {may be subjected to dynamic compression without leaking by trapping it in the inside of a hollow cone.} 
{To ensure} 
that the air volume does not leak out during an experiment{, we prepared a hollow} 
inverted cone to contain the air, and repeatedly {`ejected'} out the excess air for each given $V$. This brings the experiment close to the idealised condition of {the} Bagnold model without {suffering from an} 
air leak{, which allows us to compare the behaviour of a non-spherical bubble in a complex geometry with the simple one-dimensional Bagnold model.} Relevant situations in the real world may be found in the slamming of overturned waves where a large pocket is entrapped.\\

We use spectral analysis of the pressure time series to look for interesting resonances. In particular, we found it easy to identify the 
natural frequency {of the bubble and} 
estimate{d} its size using Minnaert's criterion. The present situation differs from Minnaert's idealised situation in that {the bubble is not spherical and in that the largest part} 
of the bubble's surface inside the cone is in contact with rigid walls. This is expected to lower the frequencies, and thereby {to lead to an overestimate of} the bubble's size{, which is confirmed by} 
our experiments. The observations of how the air pocket collapses in the presence of this tapering geometry {may} 
be useful to estimate the behaviour of entrapped air pockets in{, e.g.,} 
the slamming of waves onto corrugated surface (such as MARK III). {Our work could also give insight in}
how the approaching liquid {may invade} 
an air pocket entrapped against a corrugated wall.\\

{In comparing the initial maximal pressurisation $P_\text{broad}$ of the bubble with the one-dimensional Bagnold model, we found that the experiments were reasonably well-described by the latter, albeit for a much smaller value for the ratio of the added liquid mass and the initial bubble size ($K/x_0$) than could be expected based on the geometry of the experimental setup. This can likely be attributed to the fact that in our experiments, the linear motor needs a finite amount of time to reach the target velocity $V$, which, for a fair comparison to the Bagnold model, should happen instantly.}\\

{Our findings suggest the following experimental design that would make use of the benefits of our setup, but would bypass its flaws: One could use a similar inverted cone with a pressure sensor, partly filled with air, that is hit from above by the impact of a much heavier object, reminiscent of a pile rammer, such that it is instantly accelerated to a velocity close to that of the impactor. This would bring the setup much closer to the situation described by the Bagnold model, and would have the added benefit that it could be relatively easily scaled up, such that much smaller values of the Bagnold number $S$ (larger $1/S$) can be reached.}\\

{In our experiments we find that the measures taken to avoid air from spilling out of the impactor make it easier for liquid to enter into the air pocket, and create a pinch-off event followed by the formation of a jet (or multiple jets) that impact onto the pressure sensor and create a pressure spike} $P_{\text{peak}}${. Although these} pressures are much larger than {those associated with} the Bagnold impact{, $P_{\text{broad}}$ and} $P_{\text{max}}$, their short duration {makes them easy to separate from the rest of the signal. In addition, the fact that they are short-lived suggests} 
that they will make a {very small} 
contribution to the {total} impulse. {The} 
absence of any trend with the {velocity} $V$ indicates {a large variability caused by many small details that are not sufficiently controlled (or even controllable) in the experiments.}\\ 

{Finally, we} note that these air-pocket-mediated impact pressures are distinct from `air-cushioning' mediated impact pressures that can be found in flat-impact scenarios. In the latter case, the entrapped air layer is so thin that a sharp (near-singular) peak pressure is registered even before the trapped air layer gets enough time to evolve. This knowledge can be used to infer the shape and size of an entrapped air-layer simply from the pressure time series recorded on the impact target. In the vicinity of a sharp peak pressure, a steady pressurisation spread over time scales greater than $10^{-3}$ s (such as in figure \ref{fig:pressuretimeseries}) would imply the entrapment of a large air pocket. Conversely, a quicker (time scales shorter than $10^{-3}$)  pressurisation in the vicinity of the impact peak would suggest the entrapment of a thin air layer.\\

\section{Acknowledgements}
We acknowledge financial support from SLING (project number P14-10.1), which is partly financed by the Netherlands Organisation for Scientific Research (NWO). We also acknowledge funding via Vici grant {17070}, which is also financed by NWO.

\nocite{*}

\bibliography{apssamp}

\end{document}